\documentclass{article}

\usepackage{graphicx} \usepackage{amsmath}

\suppressfloats[t]

\begin{document} \large

\begin{center}
	{\bf Numerical simulation of the wave turbulence on the surface of a ferrofluid in a horizontal magnetic field }\\[1.0ex]
	
	\vspace{2mm}
	{\it Evgeny A. Kochurin$^1$ }\\[1.0ex]
	{$^1$Institute of Electrophysics, Ural Division,
		Russian Academy of Sciences, 106 Amundsen Street, 620016 Ekaterinburg, Russia\\
	}
\end{center}

\vspace{2mm}

\begin{abstract}
The turbulence of capillary waves on the surface of a ferrofluid with a high permeability in a horizontal magnetic field is considered in the framework of a one-dimensional weakly nonlinear model. In the limit of a strong magnetic field, the surface waves under study can propagate without distortions along or against the direction of external field, i.e., similar to Alfv\'en waves in a perfectly conducting fluid. The interaction of counter-propagating nonlinear waves leads to the  development of wave turbulence on the surface of  the liquid. The computational data show that the spectrum of turbulence is divided into two parts: a low-frequency dispersionless region, where the magnetic forces dominate and a high-frequency dispersive one, in which the influence of capillary forces becomes significant. 
In the first region, the spectrum of the surface elevation has the same exponent in $k$ and $\omega$ domains and its value is close to $-3.5$, what is in a good agreement with the estimation obtained from the dimensional analysis of the weak turbulence spectra. At the high frequencies, the computed spatial spectrum of the surface waves is close to $k^{-5/2}$ which corresponds to $\omega^{-5/3}$ in terms of the frequency. This spectrum does not coincide with the Zakharov-Filonenko spectrum obtained for pure capillary waves. A possible explanation of this fact is in the influence of coherent structures (like shock waves) usually arising in media with weak dispersion.

\end{abstract}

\section{Introduction}
Wave turbulence is observed in the interaction of nonlinear dispersive waves in many physical processes, see the review \cite{newell} and references therein. Zakharov and Filonenko \cite{Zakh} proposed a theory of weak turbulence of capillary waves on the surface of a liquid. According to which, a stationary regime of wave turbulence with an energy spectrum, now called the Zakharov-Filonenko spectrum, is formed at the boundary of a liquid. To date, the theory of weak turbulence has been very well confirmed both experimentally \cite{mezhov1, mezhov2,falcon2007, falc_exp} and numerically \cite{Push, korot,pan14,falcon14}. Physical experiments \cite{falcon2, falcon3} carried out for magnetic fluids in a magnetic field showed that the external field can modify the turbulent spectrum of capillary waves. So far, there was no theoretical explanation for this fact.

In this paper, we consider the nonlinear dynamics of the free surface of a magnetic fluid in a horizontal magnetic field. Melcher \cite{melcher1961} has shown that the problem under study is mathematically completely equivalent to the problem of the dynamics of the free surface of a dielectric fluid in a horizontal electric field. For this reason, in this work we will use the previously obtained results for non-conducting liquids in an electric field. The dynamics of coherent structures (solitons or collapses) on the surface of liquids in a magnetic (electric) field has been very well studied (see, for example, \cite{koulova18,ferro10, tao18, zu2002, gao19}). At the same time, the turbulence of surface waves in an external electromagnetic field was not theoretically investigated (except of our recent work \cite{ko_19_jetpl}). In this paper, we will show that at the free surface of a ferrofluid in a magnetic field, new wave turbulence spectra differing from the classical spectra for capillary and gravitational waves can be realized.

\section{Linear analysis}
We consider a potential flow of an ideal incompressible ferrofluid with infinite depth and a free surface in a uniform horizontal external magnetic field. The fluid is dielectric, i.e., there are no free electrical currents in the liquid. Since, the problem under consideration is anisotropic because of the existence of the separated direction of the magnetic field, we consider only plane symmetric waves propagating in the direction parallel to the external field. Let the vector of a magnetic field induction be directed along the $x$ axis (correspondingly, the $y$ axis of the Cartesian coordinate system is perpendicular to it) and has the absolute value $B$. The shape of the boundary is described by the function $y=\eta(x,t)$, for the unperturbed state $y=0$.

The dispersion relation for linear waves at the boundary of the liquid has the form \cite{melcher1961}
\begin{equation}\label{disp}\omega^2=gk+\frac{\gamma(\mu)}{\rho}B^2k^2+\frac{\sigma}{\rho}k^3,\end{equation}
where $\omega$ is the frequency, $g$ is the gravitational acceleration, $k$ is the wavenumber, $\gamma(\mu) =(\mu-1)^2 (\mu_0(\mu+1))^{-1}$ is the auxiliary coefficient, $\mu_0$ is the magnetic permeability of vacuum, $\mu$  and $\rho$ are the magnetic permeability and mass density of the liquid, respectively, and $\sigma$ is the surface tension coefficient.

Let estimate the characteristic physical scales in the problem under study. In the absence of an external field, the dispersion relation (\ref{disp}) describes the propagation of the surface gravity-capillary waves. Their minimum phase speed is determined by the formula: $v_{min}=(4 \sigma g / \rho)^{1/4}$. To obtain the characteristic magnetic field we need to equate $v_{min}^2$ to the coefficient before $k^2$ (it has a dimension of squared velocity) in the right-hand side of (\ref{disp}). Thus, the critical value of the magnetic field induction has the form 
\begin{equation}\label{field}
B_c^2=\frac{2(\rho g\sigma)^{1/2}}{\gamma(\mu)}.
\end{equation}
The characteristic scales of length and time are 
\begin{equation}\label{scale}
\lambda_0=2 \pi \left(\frac{\sigma}{g\rho}\right)^{1/2},\quad t_0=2 \pi \left(\frac{\sigma}{g^{3}\rho} \right)^{1/4}.
\end{equation}
Let us calculate the specific values of the introduced quantities for the liquid used in the experiments \cite{falcon2, falcon3}. Put the fluid parameters as follows 
$$\rho=1324\, \mbox{kg/m}^3,\quad \sigma=0.059\, \mbox{N/m}, \quad \mu=1.69.$$
Substituting these parameters into above formulas, we obtain the estimations for the characteristic quantities in the problem under study: $\lambda_0\approx 1.3$ cm, $t_0\approx0.1$ s, and $B_c\approx 196$~G. It should be noted that the critical value of the magnetic field  decreases with increasing magnetic permeability of the liquid. For the liquid with $\mu=10$, it is estimated as a relatively small quantity $B_c\approx 30$ G. Further in the work, it will be demonstrated that the magnetic wave turbulence can develop at the boundary of a ferrofluid with high magnetic permeability in the following field range: $2\leq B/B_c \leq 6$, i.e., the maximum value of $B_{max}$ used in the work is near 200 G.

Note that in the case of a strong magnetic field, the dispersion relation (\ref{disp}) must be modified taking into account the magnetization curve, as was done in \cite{zel69}. Let us rewrite the expression for the magnetization $M(H)$ of a colloidal ferrofluid composed of particles of one size \cite{rosen87}:
$$M(H)/M_{st}=\left(\coth \theta-1/\theta\right)\equiv L(\theta),\,\,\,\theta=\frac{\pi \mu_0M_d D^3H}{6 k_B T},$$
where $L(\theta)$ is the Langevin function, $D$ is the particle diameter, $M_{st}$ is the magnetic saturation, $M_d$ is the domain magnetization of the particles, $k_B$ is Boltzmann's constant, $T$ is the absolute temperature. For small values of the external field, the Langevin function can be approximated by the linear dependence $L(\theta)\approx \theta/3$. Hence, in such a situation, the fluid magnetization is linearly related with the magnetic field strength $M=\chi_i H$, where $\chi_i$ is the initial magnetic susceptibility defined as,
\begin{equation}\label{chi}\chi_i=\frac{\pi \mu_0M_d^2 d^3}{18 k_B T}.
\end{equation}
Here we took into account that $M_{st}=\phi M_d$, where $\phi$ is the volume fraction of the ferromagnetic particles in a liquid. The formula (\ref{chi}) gives a relatively large value, $\chi_i\approx 9.5$, for a fluid consisting of the  magnetite particles ($M_d\approx 4.46 \cdot 10^5$ A/m) with the characteristic size $D=22$ nm, and $T=300$~K.

We now estimate the characteristic field strength $H_0$ for which the magnetization curve becomes deviate from the linear law $\theta/3$. For the value $\theta(H_0)=1$, the relative deviation between $L(\theta(H_0))$ and $\theta(H_0)/3$ is near 6$\%$, so this equality can be used as a criterion of the characteristic field. Form the definition of $\theta$, we obtain the expression for the characteristic magnetic field strength:
\begin{equation}\label{field2}H_0=\frac{6k_BT}{\pi \mu_0 M_d D^3}.\end{equation}
For the fluid with magnetic permeability, $\mu\approx 10$, the field strength is estimated as, $H_0\approx 1.3\cdot 10^3$ A/m (we put the fluid parameters as follows: $D=22$ nm, $M_d=4.46\cdot10^5$~A/m, $T=300$ K). At the same time, the critical magnetic field defined from (\ref{field}) should be near $H_c=B_c/\mu\mu_0\approx 0.2\cdot 10^3$~A/m, which is much less than the characteristic field (\ref{field2}). The maximum value of the magnetic field used in this work can be estimated as $H_{max}=B_{max}/\mu\mu_0\approx 1.5 \cdot 10^3$~A/m, this value is close to $H_0$. Thus, for the maximum magnetic field, the magnetization curve will differ from the linear dependence. Quantitatively, this difference is estimated at around 10$\%$, which is a relatively small value. For this reason, further in the work, we will assume that the magnetization of the ferrofluid linearly depends on the magnetic field strength.

\section{Turbulence spectra for dispersionless waves}
The dispersion law (\ref{disp})  describes three types of the surface waves: gravity, capillary, and magnetic ones. The magnetic surface waves are most interesting for us in this work.  In contrast to the gravity and capillary waves, such waves propagate without dispersion. Indeed, in the following range of wavenumbers $k_{gm}\ll k\ll k_{mc}$, where $k_{gm}=g \rho /\gamma B^2$, and $k_{mc}=\gamma B^2/\sigma$,  the dispersion law (\ref{disp})  has the simple form
\begin{equation}\label{lindisp}
\omega^2=v_A^2 k^2,\qquad v_A^2=\frac{\gamma(\mu)B^2}{\rho}.
\end{equation}
The wavenumber $k_{gm}$ is transitional between the gravity and the magnetic waves, and $k_{mc}$  separates the magnetic waves from the capillary ones.

In the limit of a strong field  $B\gg B_c$, and high magnetic permeability $\mu\gg 1$, Zubarev \cite {zu2004, zuzu2006, zu2009} has found exact particular solutions of the full equations of magnetic hydrodynamics in the form of nonlinear surface waves propagating without distortions in the direction or against the direction of the external horizontal magnetic field. In fact, the solutions obtained are a complete analogy of Alfv\'en waves in a perfectly conducting fluid which can propagate without distortions along the direction of the external magnetic field. The interaction is possible only between  oppositely propagating waves, and it is elastic \cite{mhd0}. Surface waves in the high magnetic field regime studied in this work have the same properties \cite{zubkoch14}.

The classical result of studying the wave magnetohydrodynamic (MHD) turbulence is the Iroshnikov-Kraichnan spectrum \cite{irosh,kraich}. According to the phenomenological theory of Iroshnikov and Kraichnan, the turbulent spectrum for fluctuations of the local magnetic field $\delta B_k$ and the fluid velocity $\delta V_k$ has the form:
\begin{equation}\label{IK1}
|\delta B_k|^2\sim|\delta V_k|^2\sim (SV_A)^{1/2}k^{-1/2},
\end{equation}
where $V_A=(\mu_0\rho)^{-1/2}B$ is the Alvf\'en speed, $B$ is the magnetic field induction inside the fluid,  $S$ is the rate of energy dissipation per unit mass. Note that in such a model of turbulence, the fluctuations of velocity and magnetic field should be small: $\delta V_k\ll V_A$, $\delta B_k\ll B$. According to (\ref{IK1}), the turbulence spectrum for the spectral density of the system energy has the form: 
\begin{equation}\label{enIK}
\varepsilon_k \sim (S V_A) k^{-3/2}. 
\end{equation}

The spectrum (\ref{IK1}) is written in terms of fluctuations of the velocity and the magnetic field in a liquid in 3D geometry, formally we can obtain its analogue for the quantities $\eta$ and $\psi$ (the value of the velocity potential at the boundary of liquid) used in the work. To do this, let introduce the perturbations of velocity $ \delta v_k $ and magnetic field induction $\delta b_k$ at the fluid boundary $y=\eta$. From the dimensional analysis ($\delta v_k \sim k\psi_k $, $\delta b_k \sim B k \eta_k $) and the dispersion relation $\omega_k \sim k$, in the strong field limit, one can obtain the spectra:
\begin{equation}\label{IK2} |\eta_k|^2\sim|\psi_k|^2\sim k^{-5/2},\quad |\eta_\omega|^2\sim|\psi_\omega|^2\sim \omega^{-5/2}.
\end{equation}
In our recent work \cite{ko_19_jetpl}, it was observed that the slope of the spectrum for the surface elevation in $k$-space is close to $-2.5$. But the analysis of the spectrum for the quantity $\psi(k,\omega)$ was not carried out. In the present work, we will examine the realizability of the spectrum (\ref{IK2}) in detail.

The Iroshnikov-Kraichnan energy spectrum (\ref{enIK}) can be obtained with help of the dimensional analysis of the weak turbulence spectra \cite{naz2003}. The weak turbulence spectrum for the energy density of a wave system with the linear dispersion law ($\omega_k\sim k$) like (\ref{lindisp}) and quadratically nonlinearity (three-waves interactions) can be written as follows (for more details see Nazarenko's book \cite{naz2011}):
\begin{equation}\label{energy}
\varepsilon_k\sim k^{\frac{1}{2}(d-6)},\qquad k=|\textbf{k}|,
\end{equation}
where $d$ is the dimension of space. It can be seen, that the spectrum (\ref{enIK}) is a particular case of (\ref{energy}) for $d=3$. The spectrum (\ref{energy}) also describes the energy distribution of the acoustic wave turbulence \cite{zakh70,efimov18}.

The spectral  density of the system energy for our problem is related with the surface elevation spectrum as follows: $\varepsilon_k\sim \omega_k |\eta_k|^2$. Form this expression and energy spectra (\ref{energy}) we can obtain the dimensional estimations for the turbulence spectra in terms of $\eta (k,\omega)$ and $\psi (k,\omega)$:
\begin{equation}\label{sp2}
|\eta_k|^2\sim |\psi_k|^2\sim k^{-3}, \,\, |\eta_\omega|^2\sim|\psi_\omega|^2\sim \omega^{-3},\, d=2,
\end{equation}
\begin{equation}\label{sp3}
|\eta_k|^2\sim |\psi_k|^2\sim k^{-7/2}, \,\, |\eta_\omega|^2\sim|\psi_\omega|^2\sim \omega^{-7/2},\, d=1,
\end{equation}
For the  first time, the spectrum (\ref{sp2}) was obtained by Falcon in \cite{falcon3} for a normal magnetic field, in \cite{falcon2} it was shown that in a tangential field, the spectrum index shifts to the region of higher values. 

Note that dimensional estimates for the capillary turbulence spectrum can also be obtained in one-dimensional geometry. Although this derivation is formal, since the capillary waves do not satisfy the conditions of three-wave resonances in 1D geometry, it will be useful to have these estimates for the comparison with the results of our numerical simulation. Skipping the details, we write out the surface spectrum for the one-dimensional capillary wave turbulence \cite{naz2011}:
\begin{equation}\label{ZF}
|\eta_k|^2\sim k^{-17/4},\qquad |\eta_{\omega}|\sim\omega^{-19/6}.
\end{equation}
It can be seen that these relations are highly different from that obtained for pure magnetic surface waves (\ref{IK2}), (\ref{sp2}), and (\ref{sp3}). The main purpose of this work is to find out which of the spectra will be the closest to that observed in a direct numerical simulation.

\section{Results of numerical simulation}
Our numerical model is based on the weakly nonlinear approximation, when the angles of boundary inclination are small $ \alpha = | \eta_x | \ll 1 $. We consider the liquid with high magnetic permeability $ \mu \gg 1 $, i.e., the surface waves have properties similar to Alfv\'en waves.
For further analysis, it is convenient to introduce the dimensionless variables
$$\eta\to \eta \cdot \lambda_0,\quad x\to x\cdot \lambda_0,\quad t\to t \cdot t_0,\quad  \psi\to \psi \cdot \lambda_0^2/t_0,$$
where $\lambda_0$,  $t_0$ are the characteristic values of length and time (\ref{scale}). It is convenient to introduce the dimensionless parameter $\beta$ defining the magnetic field induction as follows $\beta=\sqrt{2}B/B_c$, where $B_c$ is defined by  (\ref{field}), i.e., if $B=B_c$ then $\beta^2=2$.

Below, we consider the region of magnetocapillary waves $\beta^2+k\gg 1/k$, i.e., the wavelengths for which the effect of gravitational force can be neglected. The dispersion relation (\ref{disp}) can be represented in the dimensionless form
\begin{equation}\label{disp2}\omega^2=\beta^2k^2+k^3.\end{equation}
According to (\ref{disp2}), the linear surface waves are divided into two types: low-frequency magnetic ($k\ll k_c$, $\omega\ll \omega_c$) and high-frequency capillary ($k\gg k_c$, $\omega\gg\omega_c$) waves, where $k_c$ and $\omega_c$ are the crossover wavenumber and frequency defined as
$$k_c=\beta^2,\qquad \omega_c=\sqrt{2}\beta^3.$$

The equations of the boundary motion up to the quadratically nonlinear terms were first time obtained by Zubarev in \cite{zu2004}, they can be represented in the form 
$$\psi_t=\eta_{xx}+\frac{1}{2}\left[\beta^2[(\hat k \eta)^2-(\eta_x)^2] +(\hat k \psi)^2-(\psi_x)^2\right]$$
\begin{equation}\label{eq1}+\beta^2 \left[-\hat k \eta +\hat k(\eta\hat k \eta)+\partial_x(\eta \eta_x)\right]+\hat D_k \psi,\end{equation}
\begin{equation}\label{eq2}\eta_t=\hat k \psi- \hat k(\eta \hat k \psi)-\partial_x(\eta \psi_x)+\hat D_k \eta,\end{equation}
where $\hat k$ is the integral
operator having the form: $\hat k f_k=|k| f_k$ in the Fourier representation. The operator $\hat D_k$ describes viscosity
and is defined in the $k$-space as
$$\hat D_k=-\nu (|k|-|k_d|)^2, \quad |k|\geq |k_d|;\quad  \hat  D_k=0,\quad |k|< |k_d|.$$
Here, $\nu$ is a constant, and $k_d$ is the wavenumber determining the spatial scale at which the energy dissipation occurs.

Equations (\ref{eq1}) and (\ref{eq2}) are Hamiltonian and can be derived as the variational derivatives
$$\frac{\partial \psi}{\partial t}=-\frac{\delta \mathcal{H}}{\delta \eta},\qquad \frac{\partial \eta}{\partial t}=\frac{\delta \mathcal{H}}{\delta \psi}.$$
Here,
$$\mathcal{H}=\mathcal{H}_0+\mathcal{H}_1=\frac{1}{2}\int \left[\psi \hat k \psi+\beta^2 \eta \hat k \eta+(\eta_x)^2 \right]dx-$$
$$-\frac{1}{2}\int \eta\left[(\hat k \psi)^2-(\psi_x)^2+\beta^2[(\hat k \eta)^2-(\eta_x)^2]\right]dx,$$
is the Hamiltonian of the system specifying the total energy. The terms $\mathcal{H}_0$ and $\mathcal{H}_1$ correspond to the linear and quadratic nonlinear terms in (\ref{eq1}) and (\ref{eq2}), respectively.

The spectra (\ref{IK2}), (\ref{sp2}),  and (\ref{sp3}) are obtained in the limit of an infinitely strong magnetic field, $\beta\gg1 $. It should be noted that it is not observed in formally ignoring the capillary pressure in the equations (\ref{eq1})-(\ref{eq2}). In the absence of surface tension, the interaction of counter-propagating waves results in the appearance of singular points at the boundary, at which the curvature of the surface increases infinitely \cite{kochzub18,koch18,ko_19_pjtp}.
For the realization of the regime of magnetic wave turbulence on the fluid surface, it is necessary to take into account the effect of capillary forces. It immediately follows from this fact that the weak turbulence spectra  obtained in the formal limit of a strong field will be distorted in the region of dispersive capillary waves for $k\geq k_c$.
In the current work, it will be shown that the turbulence spectrum of the surface waves is divided into two regions: a low-frequency one, at which the calculated spectrum is close to the 1D spectrum (\ref{sp3}) and a high-frequency region, in which the capillary forces deform the magnetic wave turbulence spectrum.

Let us proceed to the description of the results of our numerical experiments. To minimize the effect of coherent structures (collapses or solitons), the initial conditions for (\ref{eq1}) and  (\ref{eq2}) are taken in the form of two counter-propagating
interacting wavepackets:
\begin{equation}\label{IC}\eta_1(x)=\sum \limits_{i=1}^{4}a_i\cos(k_i x),\quad\eta_2(x)=\sum \limits_{i=1}^{4}b_i\cos(p_i x),
\end{equation}
$$\eta(x,0)=\eta_1+\eta_2,\qquad \psi(x,0)=\beta(\hat H \eta_1-\hat H \eta_2),$$
where $a_i$,  $b_i$ are the wave amplitudes (they were random), $k_i$,  $p_i$ are the wavenumbers, and $\hat H$ is the Hilbert transform defined in $k$-space as $\hat H f_k=i \mbox{sign}(k) f_k$. The spatial derivatives and integral operators were calculated using pseudo-spectral methods with the total amount of harmonics  $N$, and the time integration was performed by the fourth-order explicit Runge-Kutta method with the step $dt$. The model did not involve the mechanical pumping of the energy of the system. Hence, the average steepness of the boundary $ \overline\alpha$  was determined only by the initial conditions (\ref{IC}). The calculations were performed with the parameters $N=1024$, $dt=5\cdot 10^{-5}$, $\nu=10$, $k_d=340$. To stabilize the numerical scheme, the amplitudes of higher harmonics with $k\geq412$ were equated to zero at each step of the integration in time.

In the current work, we present the results of four numerical experiments carried out with the different values of $\beta$ and, hence, with different $B/B_c$, see the parameters used in Table~1.
From the Table~1, we can see that as the field increased, the nonlinearity level (the averaged steepness) required for the realization of a direct energy cascade decreased. Apparently, this effect is related with satisfying the conditions of three-wave resonances:
\begin{equation}\label{reson}\omega=\omega_1+\omega_2,\qquad k=k_1+k_2.\end{equation}
The conditions (\ref{reson}) are satisfied for any waves described by the linear dispersion law (\ref{lindisp}). For the waves from a high-frequency region, the capillary term in (\ref{disp2}) forbids the three-wave resonance in one-dimensional geometry. But the quasi-resonances are still possible and the probability of their realization is higher for the stronger magnetic fields. It should be noted that three-wave resonances can be achieved in 1D geometry in the absence of external field for the gravity-capillary waves near the minimum of their phase speed \cite{nonlocal15}.

The lower threshold of the parameter $\beta$ required for the turbulence development is determined by the criterion of applicability of the weakly nonlinear approximation ($\alpha\ll1$). The upper threshold of the fields is limited by the tendency to form strong discontinuities that can correspond to the appearance of vertical liquid jets \cite {kochzub18j} and the formation of a regime of strong turbulence generated by singularities \cite{kuz2004}. The range of variation of the amplitudes $a_i$,  $b_i$ in initial conditions (\ref{IC}) were chosen empirically to minimize the deviation of the model from the weakly nonlinear one. The wavenumbers $k_i$, $p_i$ were chosen in such a way that the energy exchange between nonlinear waves was the most intense. For all numerical experiments, the set of wavenumbers in (\ref{IC}) is the same and is presented in Table~2.

\begin{table}
	\caption{The parameters of four numerical experiments presented in the work, $\beta^2$ is the auxiliary dimensionless parameter, $B/B_c$ is the corresponding dimensionless magnetic field induction, $\overline\alpha$ is the averaged steepness of the surface, $\omega_c$ is the crossover frequency. }
\begin{center}
	\begin{tabular}{ccccc}
		\hline
		$\beta^2$ & 10 & 30&  50 & 70 \\
		\hline
		$B/B_c$ & 2.24 & 3.87&  5.00 & 5.92 \\
		\hline
		$\overline\alpha$ & 0.15 & 0.12& 0.10 & 0.09\\
		\hline
		$\omega_c$ & 44.72 & 232.38 & 500.00 & 828.25\\
		\hline
	\end{tabular}
\end{center}
\end{table}

\begin{table}
	\caption{The set of the wavenumbers used in the initial conditions (\ref{IC}), $i$ is the summation index, $k_i$ and $p_i$ are the wavenumbers of the wavepackets traveling to the left and to the right, respectively.}
	\begin{center}
		\begin{tabular}{ccccc}
			\hline
			i & 1& 2 & 3 & 4 \\
			\hline
			$k_i$ & 3 & 5 & 7 &9\\
			\hline
			$p_i$ & 2 & 4 & 6 &8\\
			\hline
		\end{tabular}
	\end{center}
\end{table}

Figure 1 shows the calculated energy dissipation rate $s=|d\mathcal {H}/dt|$ as a function of time for the different values of magnetic field induction. It can be seen that the system under study proceeds to the regime of quasistationary energy dissipation in times of the order $10^3 t_0$. In this mode, the probability density functions for the angles of boundary inclination become very close to Gaussian distributions (see Fig. 2). This behavior indicates the absence of strong space-time correlations and, consequently, the formation of the Kolmogorov-like spectrum of wave turbulence.
\begin{figure}
	\center{\includegraphics[width=1\linewidth]{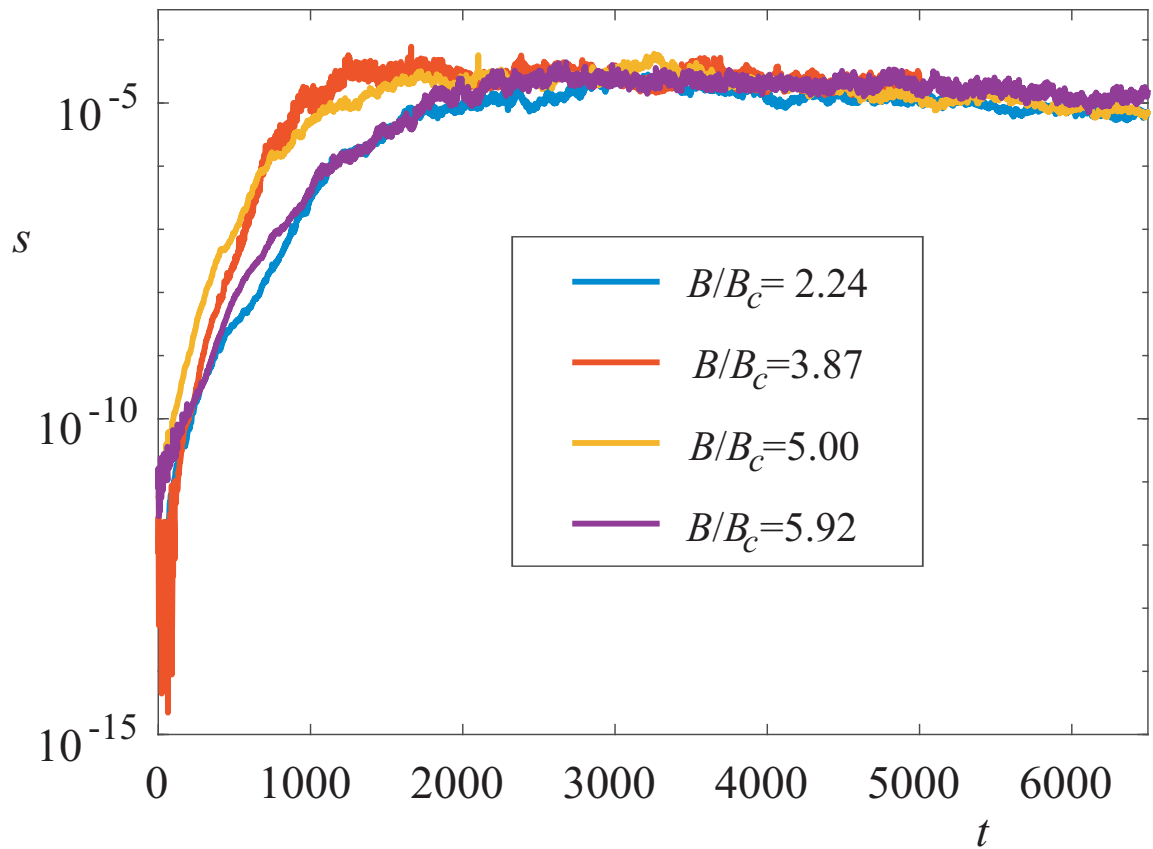}}
	\caption{\small The energy dissipation rate versus time for the different values of magnetic field.}
	\label{buble}
\end{figure}

\begin{figure}
	\center{\includegraphics[width=1\linewidth]{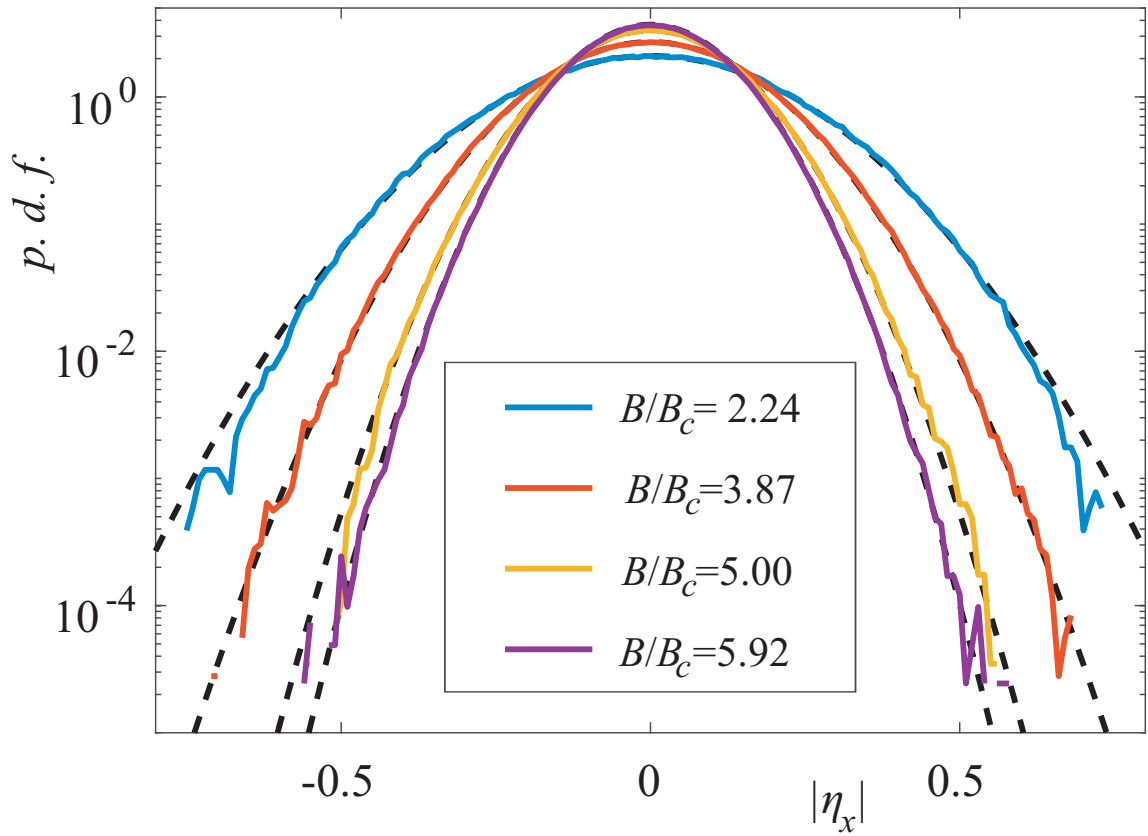}}
	\caption{\small Probability density functions (\emph{p.d.f.}) for the angles of the boundary inclination for the different values of $B/B_c$, black dotted lines correspond to Gaussian distributions.}
	\label{buble}
\end{figure}

Figure 3 shows the time-averaged spectra of the surface $I_{\eta}(k)=\overline{|\eta_k|}^2$ and the velocity potential perturbations $I_{\psi}(k)=\overline{|\psi_k|}^2$ for $B/B_c=2.24$.
As can be seen, the inertial range of the wavenumbers is splitted into two regions. In the first region with the small $k$, the spectra are in relatively good agreement with the 1D spectrum (\ref{sp3}). At the second region of high $k$, where capillarity dominates, the spectra have the different power-law:
\begin{equation}\label{IK3}I_{\eta}(k)\sim k^{-5/2}, \quad I_{\psi}(k)\sim k^{-3/2}.\end{equation}
It is interesting that the transition between two spectra occurs at higher wavenumbers than $k_c$. Since, the level of nonlinearity in this case is quite large, the observed effect can be associated with an increase in the magnetic field at steep surface inhomogeneities. Apparently, the local intensification of the magnetic field can lead to the shift of $k_c$.

\begin{figure}
	\center{\includegraphics[width=1\linewidth]{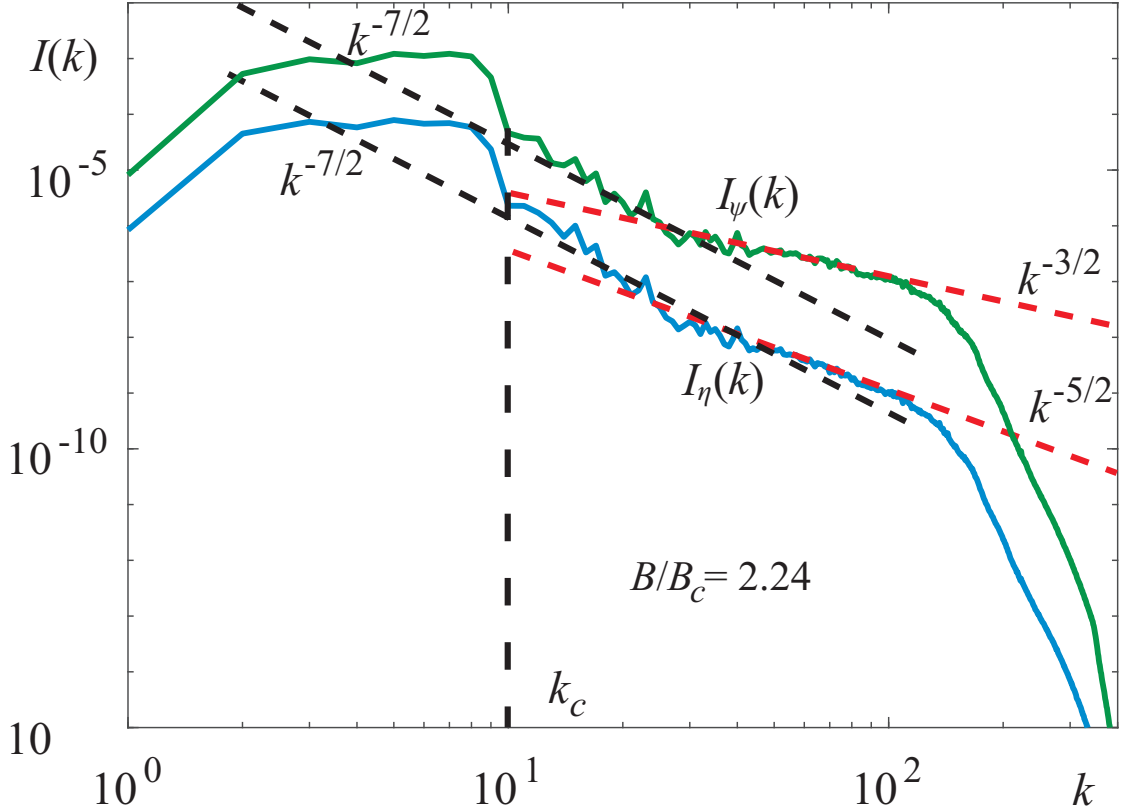}}
	\caption{\small  Time averaged spectra $I_{\eta}(k)$ (blue line) and $I_{\psi}(k)$ (green line) for $B/B_c=2.24$, the black dotted lines correspond to the 1D spectrum (\ref{sp3}), red dotted lines are the best power-law fit  (\ref{IK3}) of the calculated spectra, the vertical black dotted  line shows the crossover wavenumber $k_c$.}
	\label{buble}
\end{figure}

The spectrum (\ref{IK3}) for the surface elevation coincides with the Iroshnikov-Kraichnan one (\ref{IK2}),  but the spectrum for the velocity potential does not. Thus, the spectrum observed at high $k$ is not the MHD turbulence spectrum (\ref{IK2}), as was suggested in our previous work \cite{ko_19_jetpl}. At the same time, the spectrum (\ref{IK3}) does not coincide with the Zakharov-Filonenko spectrum (\ref{ZF}) for the pure capillary waves. Consider this spectrum in the frequency domain $\omega$. We can empirically rewrite the spectra (\ref{IK3}) in terms of $\omega$ using the dispersion relation  $ k\sim  \omega^{2/3}$ for $k\gg k_c$
\begin{equation}\label{IK4}I_{\eta}(\omega)\sim \omega^{-5/3}, \quad I_{\psi}(\omega)\sim \omega^{-1}.
\end{equation}

Figure 4 shows how the spectra $I_{\eta} (\omega)$ and $I_{\psi} (\omega) $ change as the field induction increases. From the Fig.~4~(a), it can be seen that for a relatively weak magnetic field, the spectrum is mainly determined by the relation (\ref{IK4}). As the field increases, the region of magnetic turbulence expands, see Fig.~4~(b) and (c). For the maximum magnetic field $B/B_c\approx 5.92$, the capillary waves shift to the region of viscous dissipation and almost do not have energy contribution to the spectrum of turbulence, see Fig. 4 (d). The crossover frequencies are in a good agreement with the $\omega_c$, except the first case where the level of nonlinearity is too high. In general, the calculated spectrum of turbulence in the low-frequency region is in good agreement with the spectrum (\ref{sp3}) obtained from the dimensional analysis \cite{naz2003,naz2011}.
\begin{figure}
	\center{\includegraphics[width=1\linewidth]{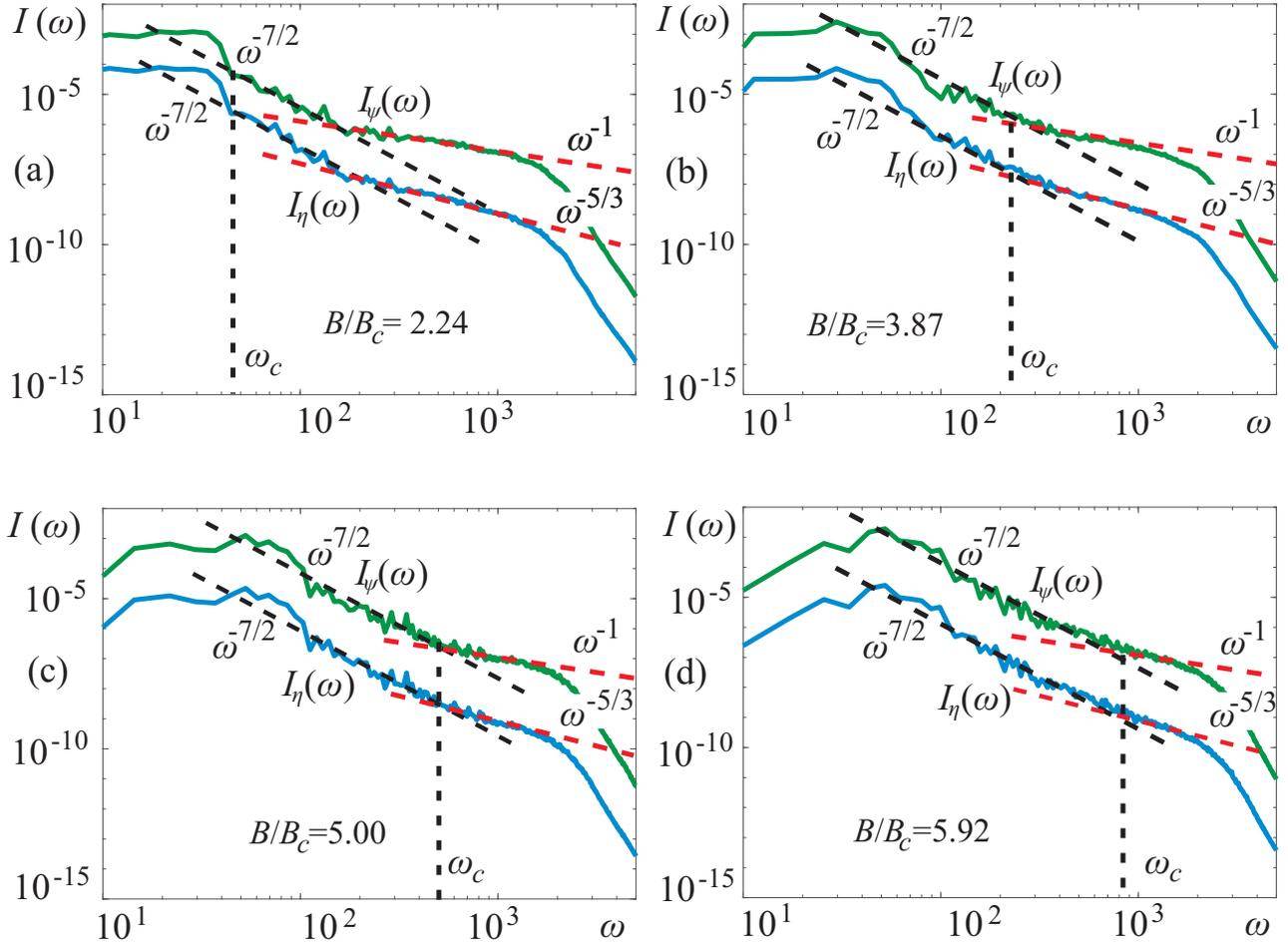}}
	\caption{\small  Time averaged spectra $I_{\eta}(\omega)$ (blue lines)  and $I_{\psi}(\omega)$ (green lines) for the different values of $B/B_c$: (a) 2.24, (b) 3.87, (c) 5.00, (d) 5.92. The black dotted lines correspond to the 1D spectrum (\ref{sp3}), red dotted lines are the best power-law fit (\ref{IK4}) of the calculated spectra, the vertical black dotted lines show the crossover frequencies $\omega_c$.}
	\label{buble}
\end{figure}

\section{Conclusion}
Thus, in the present work, a numerical study of the wave turbulence of the surface of a magnetic fluid in a horizontal magnetic field has been carried out within the framework of a one-dimensional weakly nonlinear model that takes into account the effects of capillarity and viscosity. The results show that the spectrum of turbulence is divided into two regions: a low-frequency (\emph{i}) and a high-frequency (\emph{ii}) one. 
In the region (\emph{i}), the magnetic wave turbulence is realized. The power-law spectrum of the surface elevation has the same exponent in $k$ and $\omega$ domains and is close to the value $-3.5$, which is in good agreement with the estimation (\ref{sp3}) obtained from the dimensional analysis of the weak turbulence spectra. In the high-frequency region (\emph {ii}), where the capillary forces dominate, the spatial spectrum of the surface waves is close to $k^{-5/2}$, which corresponds to $\omega^{-5/3}$ in terms of the frequency. This spectrum  does not coincide with the spectrum (\ref{ZF}) for pure capillary waves. A possible explanation of this fact is that three-wave interactions for the capillary waves are forbidden in 1D geometry and this power-law spectrum can be generated by coherent structures (like shock fronts) arising in the regime of a strong field \cite{kochzub18,koch18,ko_19_pjtp}. Its is well known that collapses and turbulence can coexist in one-dimensional models of wave turbulence \cite{MMT15,MMT17}.

In conclusion, we note that the results obtained in the work are in qualitative agreement with the experimental studies \cite{falcon2, falcon3}, in which it is shown that the external magnetic field can deform the Zakharov-Filonenko spectrum for capillary turbulence. The quantitative discrepancy may be due to the one-dimensional geometry and relatively high magnetic field leading to the nonlinear dependence of the magnetization curve, which is not taken into account in the current work.

\section*{Acknowledgments}
I am deeply grateful to N.M. Zubarev, A.I. Dyachenko, and N.B. Volkov for stimulating discussions. This work is supported by Russian Science Foundation project No. 19-71-00003.

\bibliographystyle{abbrv}
%\bibliographystyle{cas-model2-names}

%\bibliographystyle{unsrt}
\bibliography{Lit3}

\end{document}